\documentclass[conference]{IEEEtran}
\IEEEoverridecommandlockouts

\usepackage{cite}
\usepackage{amsmath,amssymb,amsfonts}
\usepackage{algorithmic}
\usepackage{graphicx}
\usepackage{textcomp}
\usepackage{xcolor}
\usepackage{acro}
\usepackage{multirow}
\usepackage{booktabs}
\usepackage{comment}
\usepackage{graphicx}  
\usepackage{subfig}    
\usepackage{caption}   
\usepackage{caption}

\usepackage{geometry}
\DeclareAcronym{ML}{
  short=ML,
  long=Machine Learning,
}
\DeclareAcronym{SVM}{
  short=SVM,
  long=Support Vector Machine,
}
\DeclareAcronym{CNN}{
  short=CNN,
  long=Convolutional Neural Network,
}
\DeclareAcronym{NIDS}{
  short=NIDS,
  long=Network Intrusion Detection System,
}
\DeclareAcronym{DNN}{
  short=DNN,
  long=Deep Neural Network,
}
\DeclareAcronym{FGSM}{
  short=FGSM,
  long=Fast Gradient Sign Method,
}
\DeclareAcronym{PGD}{
  short=PGD,
  long=Projected Gradient Descent,
}
\DeclareAcronym{IDS}{
  short=IDS,
  long=Intrusion Detection System,
}

\def\BibTeX{{\rm B\kern-.05em{\sc i\kern-.025em b}\kern-.08em
    T\kern-.1667em\lower.7ex\hbox{E}\kern-.125emX}}

\begin{document}


\title{Exploring the Effect of DNN Depth on Adversarial Attacks in Network Intrusion Detection Systems \\ \small Authors’ draft for soliciting feedback: October 22, 2025}

\author{\IEEEauthorblockN{Mohamed elShehaby}
\IEEEauthorblockA{\textit{Systems and Computer Engineering} \\
\textit{Carleton University}\\
Ottawa, Canada \\
mohamedelshehaby@cmail.carleton.ca}
\and
\IEEEauthorblockN{Ashraf Matrawy}
\IEEEauthorblockA{\textit{School of Information Technology} \\
\textit{Carleton University}\\
Ottawa, Canada \\
Ashraf.Matrawy@carleton.ca}
}

\maketitle

\begin{abstract}
Adversarial attacks pose significant challenges to \ac{ML} systems and especially Deep Neural Networks (DNNs) by subtly manipulating inputs to induce incorrect predictions. This paper investigates whether increasing the layer depth of deep neural networks affects their robustness against adversarial attacks in the \ac{NIDS} domain. We compare the adversarial robustness of various deep neural networks across both \ac{NIDS} and computer vision domains (the latter being widely used in adversarial attack experiments). Our experimental results reveal that in the NIDS domain, adding more layers does not necessarily improve their performance, yet it may actually significantly degrade their robustness against adversarial attacks. Conversely, in the computer vision domain, adding more layers exhibits a more modest impact on robustness. These findings can guide the development of robust neural networks for \ac{NIDS} applications and highlight the unique characteristics of network security domains within the \ac{ML} landscape.

\end{abstract}


\section{Introduction}

Machine Learning (ML) applications have become widespread across numerous domains, with network security being no exception. The integration of \ac{ML} techniques, particularly Deep Neural Networks (DNNs), into Network Intrusion Detection Systems (NIDS) has demonstrated significant improvements in detecting and mitigating cyber threats \cite{shehaby2023adversarial2}. However, recent research has highlighted critical vulnerabilities in these deep learning systems, particularly their susceptibility to adversarial attacks \cite{SHE024introducing}.

Adversarial attacks represent a sophisticated class of threats that subtly manipulate input data to deceive \ac{ML} models, causing them to produce incorrect predictions while appearing legitimate to human observers \cite{goodfellow2014explaining}. These attacks can severely compromise the accuracy, robustness, trustworthiness, and overall security of \ac{ML}-based systems \cite{ibitoye2025threat}. In the context of network security, such vulnerabilities could lead to the evasion of intrusion detection mechanisms, potentially allowing malicious activities to go undetected.


The motivation for this research stems from the observed relationship between neural network depth and model accuracy. Deep neural networks typically outperform shallow networks due to their ability to learn abstract and hierarchical representations through multiple layers of nonlinear transformations \cite{mhaskar2016deep}. However, while the relationship between depth and accuracy is well-established, the relationship between depth and robustness against adversarial attacks remains underexplored in NIDS domain. This gap raises a fundamental question: in the NIDS domain, does the improved representational capacity of deeper networks translate to enhanced robustness against adversarial attacks, or does increased depth introduce new vulnerabilities?

This paper addresses this question by conducting a comprehensive empirical study examining the relationship between layer depth and adversarial robustness in deep neural networks. We evaluate this relationship in the network intrusion detection domain and compare it to the computer vision domain to identify the unique characteristics of network security domains within the ML landscape.



\begin{figure}[]
\centering
\includegraphics[width=1\linewidth,keepaspectratio=true]{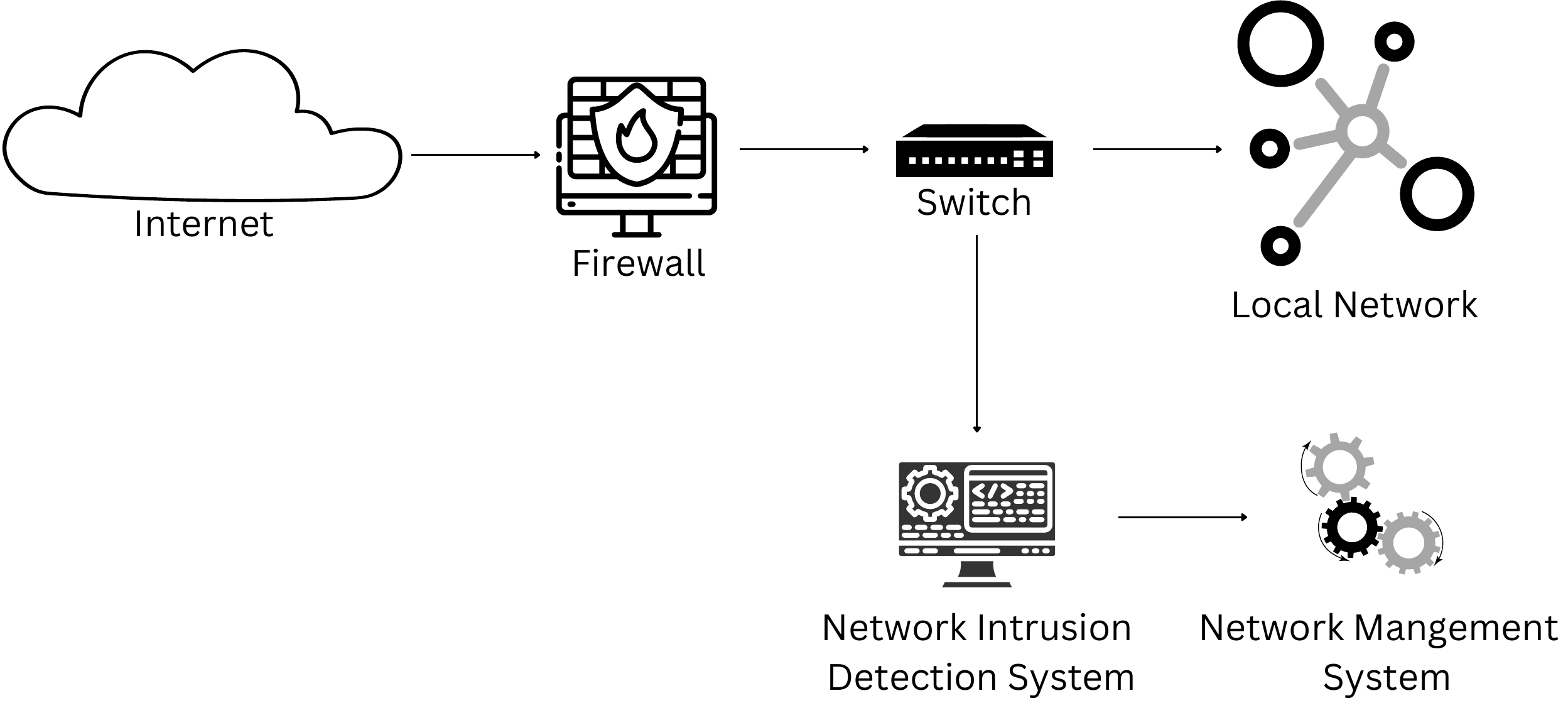}
\caption{Network Intrusion Detection System's Deployment \cite{shehaby2023adversarial2}}
\label{fig:NIDS}
\end{figure}

Our contributions are threefold; \textbf{First}, we conduct a systematic empirical analysis of how layer depth affects adversarial robustness in deep neural networks within the \ac{NIDS} domain using the CSE-CIC-IDS2018 dataset \cite{sharafaldin2018toward}. \textbf{Second}, we perform a comparative study of adversarial robustness across \ac{NIDS} and computer vision domains using MNIST \cite{lecun2002gradient}, demonstrating fundamental differences in how depth affects robustness across domains due to distinct data characteristics. \textbf{Third}, we explore and analyze the underlying reasons for our findings, examining the fundamental differences between network traffic data and image data that contribute to their distinct adversarial robustness behaviors.



\section{Related Work}
\label{related}
Numerous researchers have studied the relationship between neural network architecture and adversarial robustness, including the effects of activation functions \cite{salimi2023learning} and skip connections \cite{cazenavette2021architectural}. Regarding the relationship between neural network size and adversarial robustness, Huang et al. \cite{huang2021exploring} highlighted that increasing model parameters does not necessarily lead to more adversarially robust models. Simon-Gabriel et al. \cite{simon2019first} established theoretically that adversarial vulnerability scales with input dimensionality. However, none of these studies focused on DNN's depth in the NIDS domain, which presents unique characteristics such as lower-dimensional structured features and different data distributions compared to computer vision datasets. Moreover, through systematic comparison between NIDS and computer vision domains in terms of depth-robustness relationships, we combine the study of neural network parameters and dimensionality to account for the fundamental differences between these domains.

\section{Background}

\subsection{ML-based Network Intrusion Detection Systems}

Network intrusion detection systems identify network threats and alert management of potential breaches, as shown in Figure \ref{fig:NIDS}. Signature-based detection uses predetermined attack patterns, achieving low false positive rates but missing unknown attacks \cite{kumar2012signature}, while anomaly-based detection identifies deviations from normal activity, detecting unknown threats but often producing high false positive rates \cite{chandola2009anomaly}. Machine learning approaches have demonstrated high accuracy in addressing these challenges \cite{ahmad2021network}, yet remain vulnerable to adversarial attacks \cite{papernot2018sok}.
\begin{figure}[]
\centering
\includegraphics[width=1\linewidth,keepaspectratio=true]{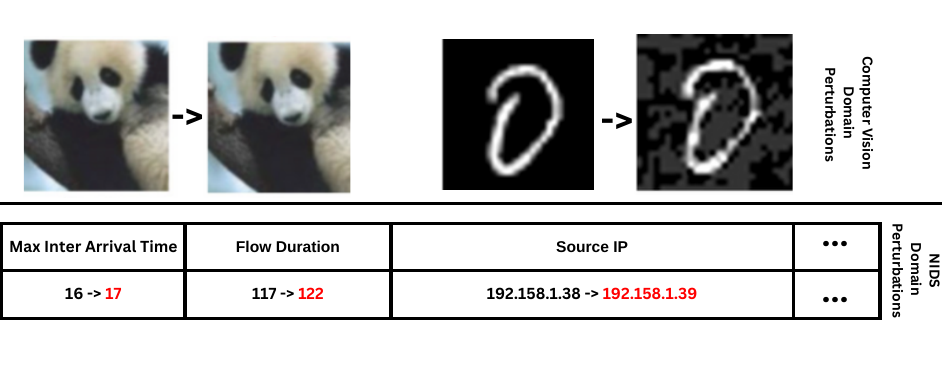}
\caption{Comparison of adversarial perturbations in computer vision (top) and NIDS (bottom) domains, illustrating the constrained nature of network traffic features versus continuous pixel modifications. Adding adversarial perturbations to image pixels produces a visually similar picture that preserves semantic content, whereas adding perturbations to network traffic flows in the NIDS domain can alter the network and/or malicious functionality of the flow.}
\label{fig:PerNIDSvsImg}
\end{figure}


\subsection{Adversarial Attacks}

Adversarial attacks against \ac{ML} models were first systematically studied in the computer vision domain. These attacks involve adding carefully crafted perturbations to input data that are imperceptible to humans but can cause significant misclassification by \ac{ML} models. 

A cornerstone and one of the most widely used evasion adversarial attack methods is the Fast Gradient Sign Method (FGSM) \cite{goodfellow2014explaining}, proposed by Goodfellow et al. FGSM generates adversarial examples by adding perturbations in the direction of the gradient of the loss function with respect to the input. The adversarial example is computed as:

\begin{equation}
x_{adv} = x + \epsilon \cdot \text{sign}(\nabla_x J(\theta, x, y))
\label{eq:fgsm}
\end{equation}

where \(x\) is the original input, \(x_{adv}\) is the adversarial example, \(\epsilon\) is the perturbation magnitude, \(\nabla_x J(\theta, x, y)\) is the gradient of the loss function \(J\) with respect to the input \(x\), \(\theta\) represents the model parameters, and \(y\) is the true label. The sign function extracts only the direction of the gradient, ensuring that each feature is perturbed by exactly \(\epsilon\) in the direction that maximally increases the loss. The epsilon parameter controls the magnitude of the perturbation and represents a trade-off between attack effectiveness and perturbation visibility. Larger epsilon values create more effective attacks but produce more noticeable perturbations, while smaller values maintain imperceptibility but may fail to fool the model. In the context of \ac{NIDS}, epsilon determines how much each network traffic feature can be modified while maintaining the attack's stealthiness.

Adversarial attacks in network intrusion detection systems face fundamentally different constraints compared to computer vision applications, as seen in Figure \ref{fig:PerNIDSvsImg},. In computer vision, perturbations are applied to continuous pixel values within a fixed range, allowing attackers to make subtle, imperceptible changes across the entire image space. 

However, NIDS adversarial attacks must operate within strict constraints where features have specific numerical bounds, categorical relationships, and functional dependencies that must be preserved to maintain valid network traffic \cite{11160917}. In a nutshell, modifying features in NIDS domain requires respecting network protocols and temporal relationships, whereas computer vision attacks only need to maintain visual imperceptibility \cite{10811295}.

\section{Experimental System Description and Methodology}
\begin{figure}[]
\centering
\subfloat[Dataset's Continuous Features' Ranges]{%
    \includegraphics[width=\linewidth,height=0.15\textheight]{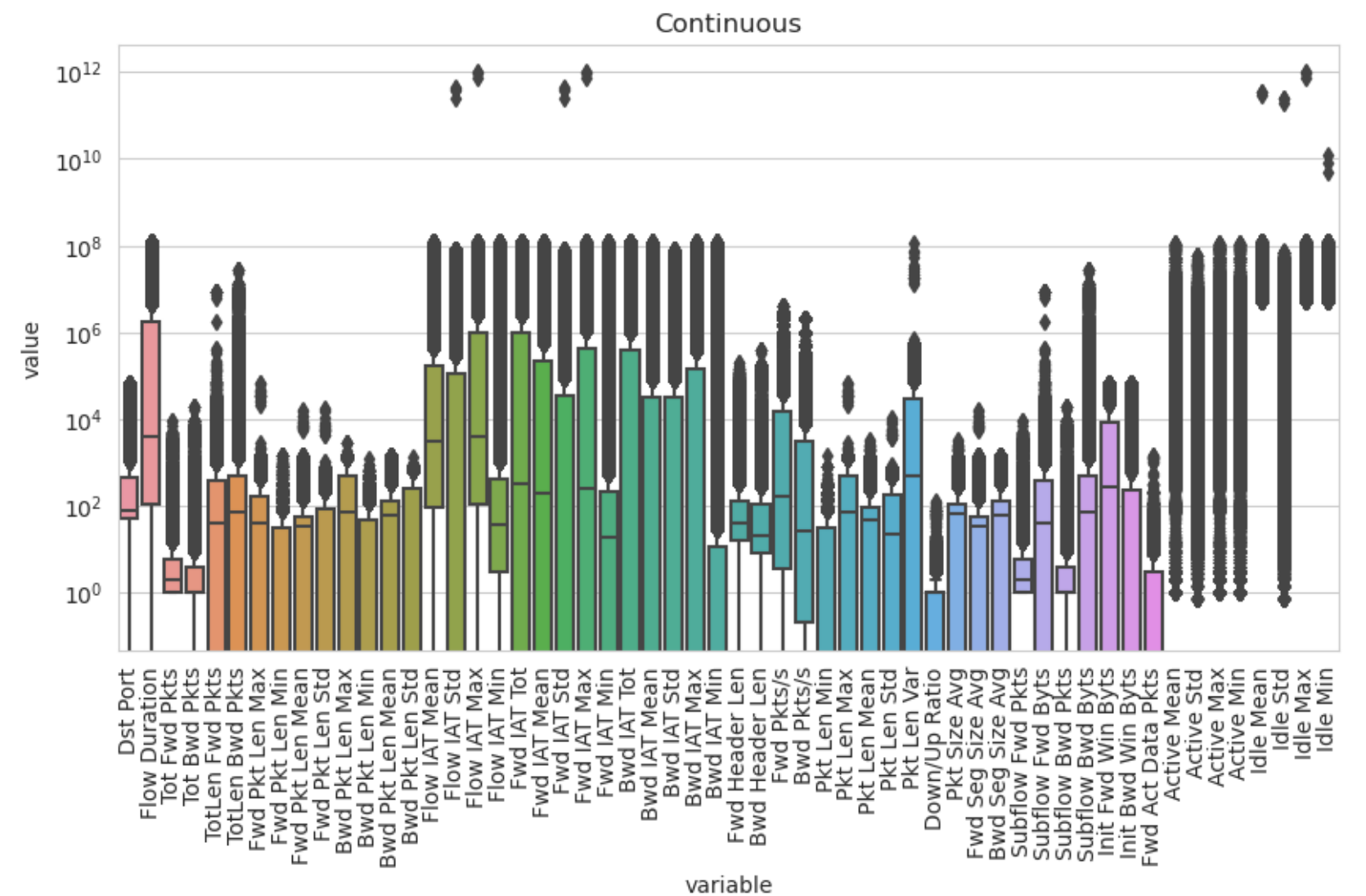}%
    \label{fig:ContBefore}%
}

\subfloat[Dataset's Continuous Features' Ranges after Standardization]{%
    \includegraphics[width=\linewidth,height=0.15\textheight]{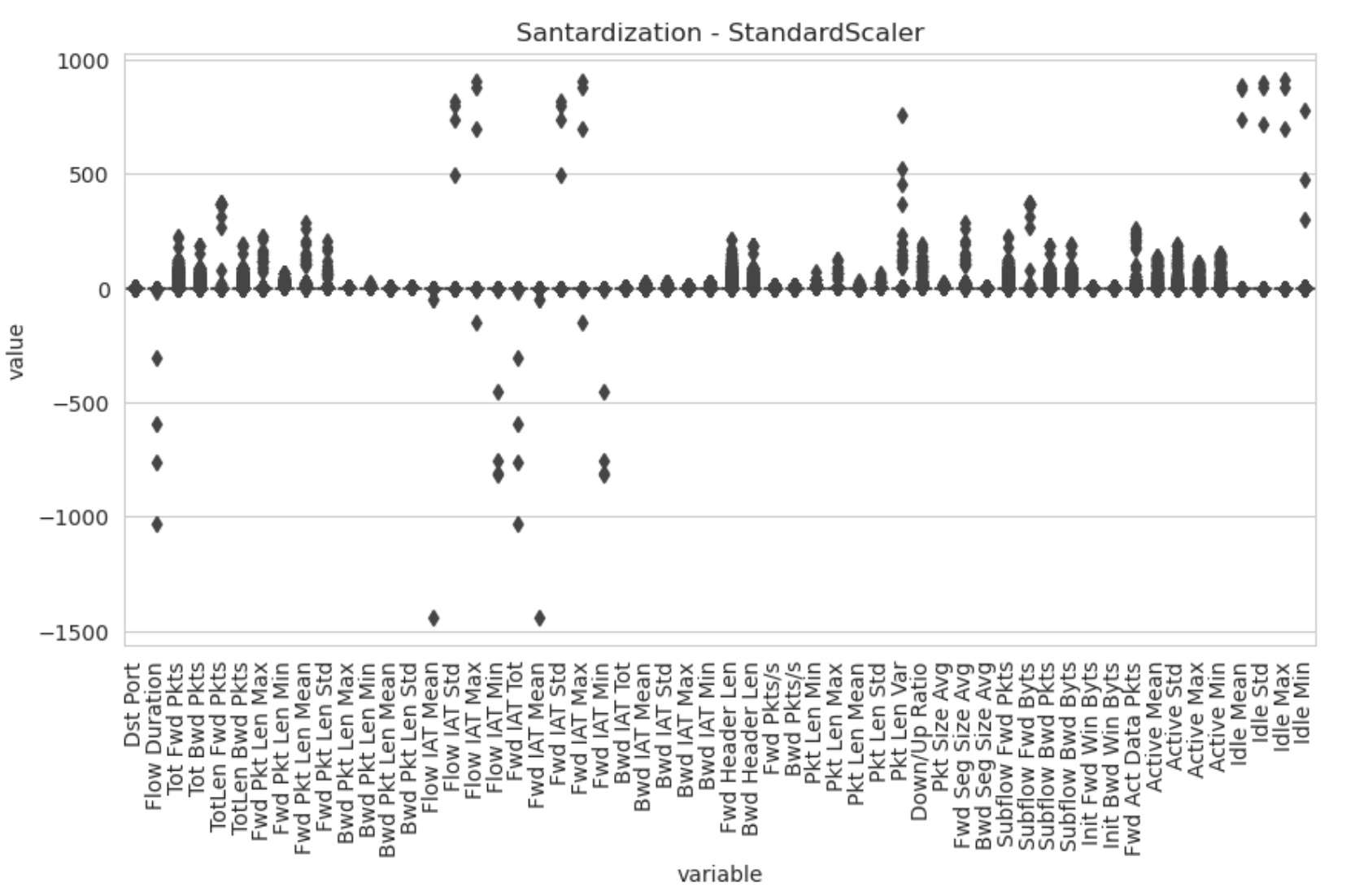}%
    \label{fig:ContAfter}%
}
\caption{Comparison of continuous features of CSE-CIC-IDS2018 before and after standardization. (a) Original features exhibit extreme range variations with numerous outliers. (b) Standardized features are centered at zero with unit variance, reducing the overall scale to [-1000, 1000] while maintaining the relative structure and distributional characteristics of each feature, ensuring equal contribution of all features during model training.}

\label{fig:continuous_features}
\end{figure}
\subsection{Datasets and Preprocessing}


For our experiments, we utilized the CSE-CIC-IDS2018 dataset \cite{sharafaldin2018toward} for the NIDS domain and MNIST \cite{lecun2002gradient} for computer vision. The CSE-CIC-IDS2018 dataset is a comprehensive benchmark for network intrusion detection, containing a wide range of real-world benign and malicious network traffic features collected from simulated enterprise environments. In contrast, the MNIST dataset is a standard benchmark in computer vision, consisting of 70,000 images of handwritten digits (0–9), each represented as a 28×28 pixel grid, and is widely used for evaluating image classification models.

Our CSE-CIC-IDS2018 dataset preprocessing involved data cleansing to remove features with null values, infinite values, NaNs, or single values, and converting string-formatted numbers to numeric values. We identified and excluded outlier samples that could disrupt model training and statistical analysis \cite{ferrari2020dealing}. To accomplish this, we applied a probability-based method to remove extreme values from all numerical features. The dataset was then standardized, as seen in Figure \ref{fig:continuous_features}, and one-hot encoded. We performed correlation analysis to eliminate highly correlated features and used feature importance ranking with Random Forest to select the most relevant features \cite{louppe2013understanding}. After preprocessing, the dataset had a 78-feature vector. 

As for The MNIST dataset, it was preprocessed by normalizing pixel values from the original range to to ensure numerical stability during training. The original training set was split into 50,000 training samples and 10,000 validation samples, while the test set remained at 10,000 samples. Each 28×28 grayscale image was flattened into a 784-dimensional feature vector to serve as input to the fully connected neural networks.

\subsection{Experimental Neural Networks Architectures}

\begin{table}[]
\centering
\setlength{\tabcolsep}{15pt}
\begin{tabular}{llc}
\hline
Model & Architecture & Hidden Layers \\
\hline
Model 1 & $n$-64-$k$ & 1  \\
Model 2 & $n$-64-128-$k$ & 2  \\
Model 3 & $n$-64-128-128-$k$ & 3  \\
Model 4 & $n$-64-128-512-128-$k$ & 4  \\
Model 5 & $n$-64-128-512-128-64-$k$ & 5 \\
\hline
\end{tabular}
\caption{Neural network architectures where $n$ represents the input dimension (784 for MNIST, 78 for CSE-CIC-IDS2018) and $k$ represents the output dimension (10 for MNIST multi-class classification, 1 for CSE-CIC-IDS2018 binary classification).}
\label{tab:model_architectures}
\end{table}
\begin{figure}[!t]
\centering
\includegraphics[width=\linewidth,keepaspectratio=true]{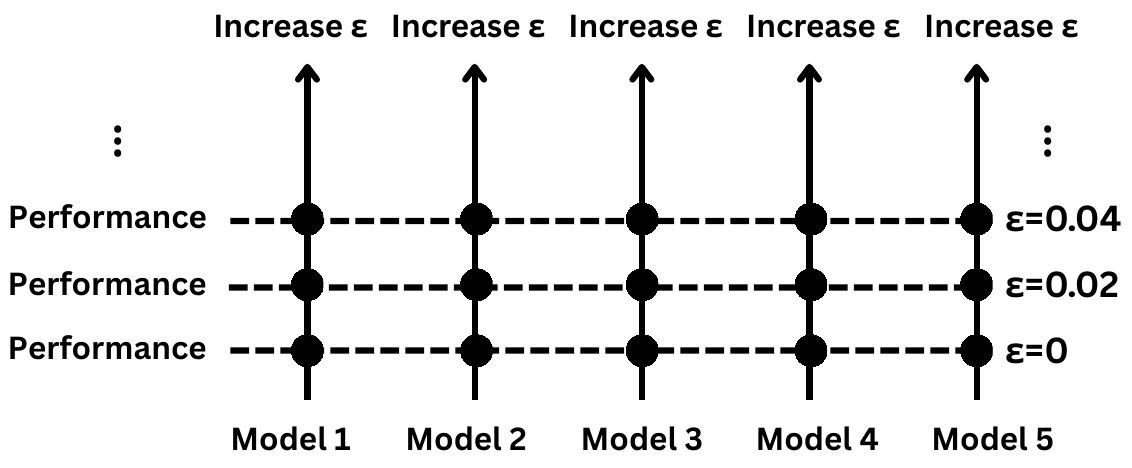}
\caption{Experimental Scenario}
\label{fig:ExpScen}
\end{figure}

To investigate the relationship between network depth and adversarial robustness, we designed five fully-connected neural network architectures with progressively increasing complexity. All models share a common structure: an input layer with $n$ neurons (where $n=784$ for MNIST and $n=78$ for CSE-CIC-IDS2018), followed by one to five hidden layers, and an output layer with $k$ neurons (where $k=10$ for MNIST multi-class classification and $k=1$ for CSE-CIC-IDS2018 binary classification). Model 1 serves as the baseline with a single hidden layer of 64 neurons (architecture: $n$-64-$k$). Models 2 through 5 systematically increase in depth: Model 2 adds a second hidden layer of 128 neurons ($n$-64-128-$k$), Model 3 adds another 128 neurons hidden layer ($n$-64-128-128-$k$), Model 4 introduces four hidden layers with widths of 64, 128, 512, and 128 neurons ($n$-64-128-512-128-$k$), and Model 5 represents the deepest architecture with five hidden layers of 64, 128, 512, 128, and 64 neurons ($n$-64-128-512-128-64-$k$).  

\subsection{Experimental Scenario}

Our experimental approach follows a systematic methodology illustrated in Figure \ref{fig:ExpScen}. The experimental scenario involves training multiple model variants, evaluating their performance on clean data, generating adversarial attacks, and assessing robustness under adversarial conditions.

Specifically, we train five neural network models of increasing depth for each domain, CSE-CIC-IDS2018 for network intrusion detection and MNIST for computer vision. For each trained model, we first evaluate its performance (precision, recall, F1-score, and accuracy) on unperturbed (clean) test data. We then generate adversarial examples using the Fast Gradient Sign Method (FGSM), systematically varying the perturbation parameter $\varepsilon$, as defined in Equation~\eqref{eq:fgsm}.

When $\varepsilon=0$ no perturbation is added to the input, meaning there is no adversarial attack and the model is tested on clean data. We gradually increase $\varepsilon$ to introduce stronger adversarial perturbations, evaluating each model’s performance at every step. This process is repeated for all model depths in both domains, allowing us to analyze how increasing network depth and adversarial strength (as controlled by $\varepsilon$) affect model robustness. By comparing results across both the NIDS and computer vision domains, we aim to uncover domain-specific patterns in adversarial vulnerability and the impact of network depth.

\begin{table}[]

\centering

\begin{tabular}{lcccccc}
\hline
  $\varepsilon=$ & 0.0 & 0.02 & 0.04 & 0.06 & 0.08 & 0.1  \\
\hline
\multicolumn{7}{c}{\textbf{Model 1 (1 hidden layer)}} \\
\hline
Precision & 0.994 & 0.995 & 0.995 & 0.995 & 0.996 & 0.996 \\
Recall & 1.000 & 1.000 & 0.999 & 0.998 & 0.933 & 0.700 \\
F1-Score & 0.997 & 0.997 & 0.997 & 0.996 & 0.963 & 0.822 \\
Accuracy & 0.999 & 0.999 & 0.999 & 0.999 & 0.986 & 0.939 \\
\hline
\multicolumn{7}{c}{\textbf{Model 2 (2 hidden layers)}} \\
\hline
Precision & 0.995 & 0.996 & 0.996 & 0.997 & 0.997 & 0.998 \\
Recall & 1.000 & 0.999 & 0.999 & 0.978 & 0.885 & 0.533 \\
F1-Score & 0.997 & 0.997 & 0.997 & 0.987 & 0.938 & 0.695 \\
Accuracy & 0.999 & 0.999 & 0.999 & 0.995 & 0.977 & 0.906 \\
\hline
\multicolumn{7}{c}{\textbf{Model 3 (3 hidden layers)}} \\
\hline
Precision & 0.996 & 0.996 & 0.996 & 0.996 & 0.994 & 1.000 \\
Recall & 0.999 & 0.999 & 0.999 & 0.956 & 0.431 & 0.285 \\
F1-Score & 0.997 & 0.997 & 0.997 & 0.976 & 0.601 & 0.443 \\
Accuracy & 0.999 & 0.999 & 0.999 & 0.990 & 0.886 & 0.857 \\
\hline
\multicolumn{7}{c}{\textbf{Model 4 (4 hidden layers)}} \\
\hline
Precision & 0.996 & 0.996 & 0.996 & 0.996 & 0.998 & 1.000 \\
Recall & 1.000 & 0.999 & 0.999 & 0.954 & 0.539 & 0.325 \\
F1-Score & 0.998 & 0.997 & 0.997 & 0.974 & 0.700 & 0.490 \\
Accuracy & 0.999 & 0.999 & 0.999 & 0.990 & 0.908 & 0.865 \\
\hline
\multicolumn{7}{c}{\textbf{Model 5 (5 hidden layers)}} \\
\hline
Precision & 0.996 & 0.996 & 0.997 & 0.998 & 1.000 & 0.000 \\
Recall & 0.999 & 0.999 & 0.915 & 0.592 & 0.246 & 0.000 \\
F1-Score & 0.997 & 0.997 & 0.954 & 0.743 & 0.395 & 0.000 \\
Accuracy & 0.999 & 0.999 & 0.982 & 0.918 & 0.849 & 0.800 \\
\hline
\end{tabular}
\caption{CSE-CIC-IDS2018: Performance metrics for all models under FGSM attack}
\label{tab:CIC_model_metrics}
\end{table}

\subsection{Attack Space in Our Experiments on NIDS Dataset}
Problem-space adversarial attacks involve modifications to the original input (network traffic) intended to create perturbations in specific features within the feature vector after feature extraction. In other words, problem-space modifications map to feature-space perturbations, a process referred to as inverse feature-mapping \cite{pierazzi2020intriguing}. Every problem-space evasion attack naturally will eventually lead to a feature-space attack. Therefore, testing the effect of adversarial attacks on neural networks in the feature space inherently evaluates their robustness against problem-space attacks. For this reason, our experiments focus on the feature-space to assess the inherent robustness of different network depths.

\begin{table}[]

\centering

\begin{tabular}{lcccccc}
\hline
  $\varepsilon=$ & 0.0 & 0.02 & 0.04 & 0.06 & 0.08 & 0.1  \\
\hline
\multicolumn{7}{c}{\textbf{Model 1 (1 hidden layer)}} \\
\hline
Precision & 0.884 & 0.845 & 0.765 & 0.633 & 0.508 & 0.395 \\
Recall & 0.883 & 0.845 & 0.764 & 0.624 & 0.476 & 0.353 \\
F1-Score & 0.883 & 0.845 & 0.765 & 0.628 & 0.492 & 0.373 \\
Accuracy & 0.883 & 0.845 & 0.764 & 0.624 & 0.476 & 0.353 \\
\hline
\multicolumn{7}{c}{\textbf{Model 2 (2 hidden layers)}} \\
\hline
Precision & 0.888 & 0.841 & 0.763 & 0.649 & 0.543 & 0.456 \\
Recall & 0.889 & 0.843 & 0.765 & 0.649 & 0.530 & 0.425 \\
F1-Score & 0.889 & 0.842 & 0.764 & 0.649 & 0.537 & 0.440 \\
Accuracy & 0.889 & 0.843 & 0.765 & 0.649 & 0.530 & 0.425 \\
\hline
\multicolumn{7}{c}{\textbf{Model 3 (3 hidden layers)}} \\
\hline
Precision & 0.874 & 0.833 & 0.743 & 0.626 & 0.545 & 0.497 \\
Recall & 0.874 & 0.833 & 0.749 & 0.639 & 0.544 & 0.479 \\
F1-Score & 0.874 & 0.833 & 0.746 & 0.633 & 0.545 & 0.488 \\
Accuracy & 0.874 & 0.833 & 0.749 & 0.639 & 0.544 & 0.479 \\
\hline
\multicolumn{7}{c}{\textbf{Model 4 (4 hidden layers)}} \\
\hline
Precision & 0.906 & 0.841 & 0.748 & 0.632 & 0.567 & 0.513 \\
Recall & 0.907 & 0.842 & 0.750 & 0.642 & 0.565 & 0.501 \\
F1-Score & 0.907 & 0.842 & 0.749 & 0.637 & 0.566 & 0.507 \\
Accuracy & 0.907 & 0.842 & 0.750 & 0.642 & 0.565 & 0.501 \\
\hline
\multicolumn{7}{c}{\textbf{Model 5 (5 hidden layers)}} \\
\hline
Precision & 0.894 & 0.850 & 0.727 & 0.625 & 0.533 & 0.465 \\
Recall & 0.895 & 0.850 & 0.734 & 0.632 & 0.535 & 0.451 \\
F1-Score & 0.894 & 0.850 & 0.731 & 0.628 & 0.534 & 0.458 \\
Accuracy & 0.895 & 0.850 & 0.734 & 0.632 & 0.535 & 0.451 \\
\hline
\end{tabular}
\caption{MNIST: Performance metrics for all models under FGSM attack}
\label{tab:mnist_model_metrics}
\end{table}

\section{Experimental Results and Analysis}

\begin{figure*}[]
\centering
\includegraphics[width=\linewidth,height=0.32\textheight]{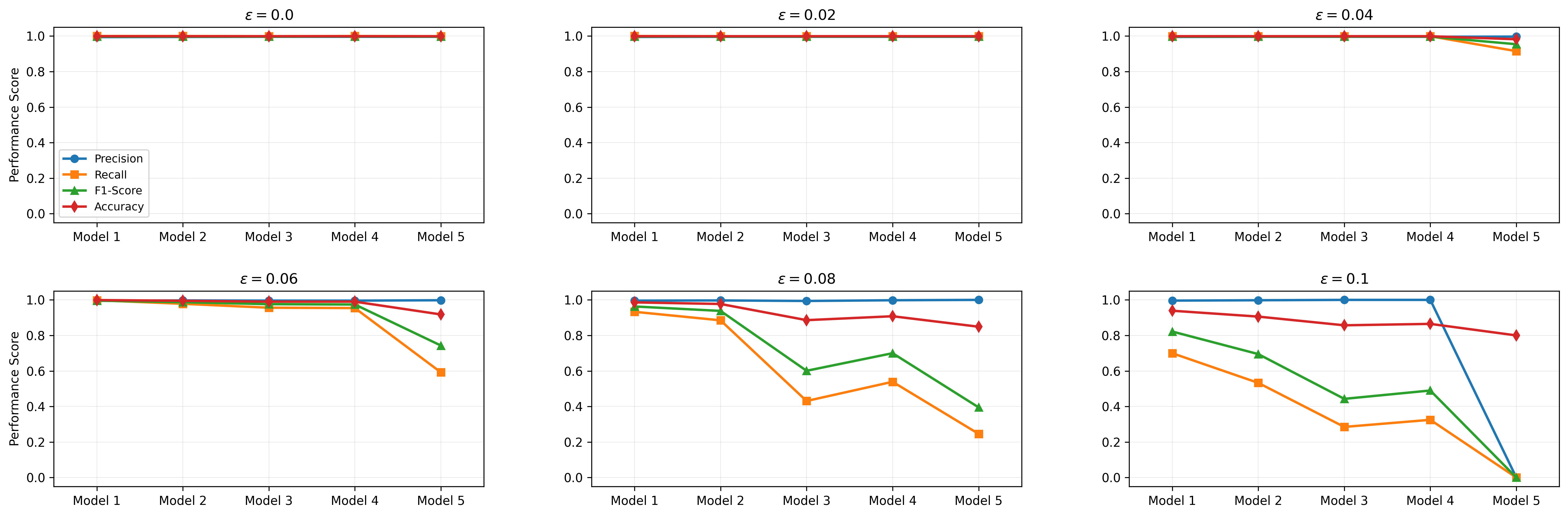}
\caption{Performance vs Network Depth under FGSM Attack at Different Epsilon Values on CSE-CIC-IDS2018 Dataset}
\label{fig:res1}
\end{figure*}

\begin{figure*}[]
\centering
\includegraphics[width=\linewidth,height=0.32\textheight]
{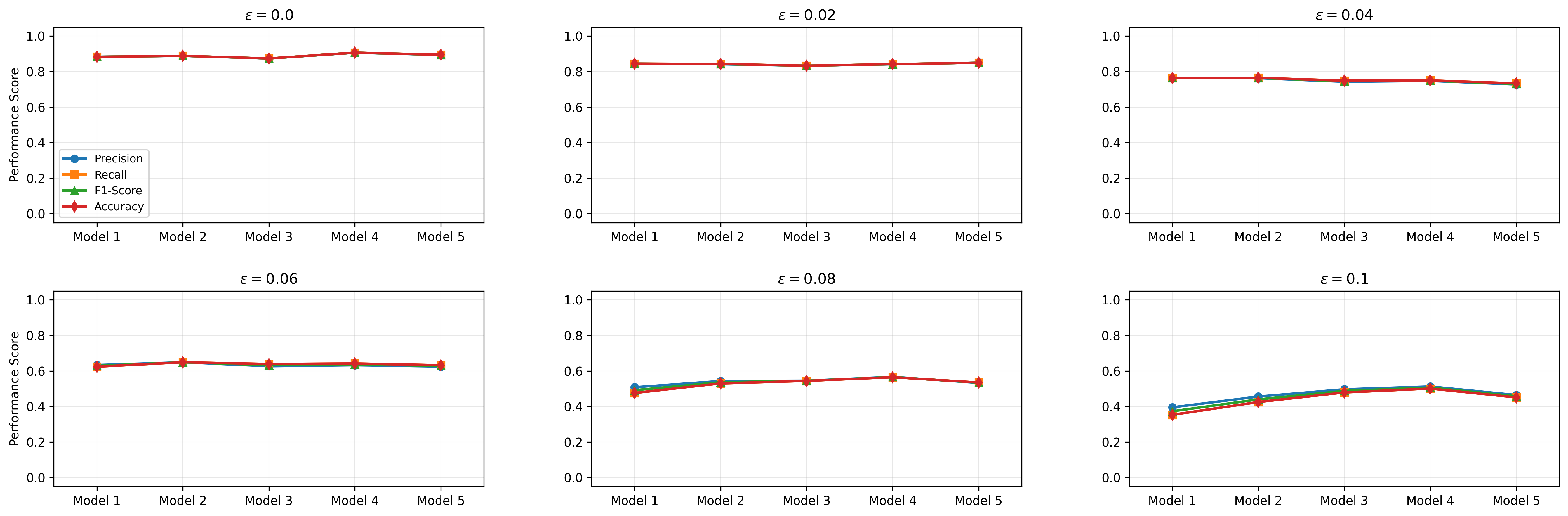}
\caption{Performance vs Network Depth under FGSM Attack at Different Epsilon Values on MNIST}
\label{fig:res2}
\end{figure*}

\subsection{Performance in NIDS Domain}
Table \ref{tab:CIC_model_metrics} and Figure \ref{fig:res1} highlight the impact of increasing network depth and FGSM attack strength on model performance for the CSE-CIC-IDS2018 dataset. When \(\varepsilon=0\) (no attack), all models achieve high accuracy, precision, recall, and F1-score, demonstrating that in NIDS domain adding more layers does not necessarily improve performance. However, as \(\varepsilon\) increases, deeper networks become more vulnerable, with their performance degrading more quickly than shallower models. This demonstrates that greater depth can actually amplify the effect of adversarial attacks in the NIDS domain, emphasizing the need for caution when designing deeper architectures for adversarial robustness. It is worth noting that even at higher \(\varepsilon\) values, shallower models in the NIDS domain exhibited high resilience to adversarial attacks.


\subsection{Performance in Computer Vision Domain}

Table \ref{tab:mnist_model_metrics} and Figure \ref{fig:res2} present the results of FGSM attacks at different \(\varepsilon\) levels on MNIST models with varying depths. At \(\varepsilon=0\), models exhibit good performance across all metrics. As \(\varepsilon\) increases, the adversarial effect reduces precision, recall, F1-score, and accuracy; however, unlike the CSE-CIC-IDS2018 results, deeper MNIST models generally maintain more consistent performance. Therefore, in computer vision, deeper neural networks do not have a huge effect on the effectiveness of FGSM, and the main effect comes from \(\varepsilon\) values. 

\subsection{Results' Analysis}

\begin{itemize}

  \item The contrasting behavior observed between MNIST and CSE-CIC-IDS2018 datasets regarding network depth and adversarial robustness can be explained by fundamental differences in their input dimensionality and data characteristics. 

  \item In image processing (MNIST), the input vector is large, so even shallow models are highly susceptible to adversarial attacks; here, the most influential parameter for robustness is $\varepsilon$, not network depth. Adding more layers does not significantly change robustness, as even shallow networks are complex enough to be vulnerable due to the high input dimensionality.

 
\item On the other hand, the feature space in the NIDS domain (CSE-CIC-IDS2018) is smaller, meaning shallow models naturally act as a line of defense against adversarial attacks, often showing strong resilience even at higher \(\varepsilon\) values due to the limited attack space. This is why defenses like model distillation and feature reduction are sometimes effective for network security, as they further constrain adversarial opportunities \cite{shehaby2023adversarial2}. However, increasing the number of parameters and layers rapidly expands the attack surface, making deeper NIDS models more vulnerable to adversarial perturbations.

\item From a theoretical perspective, the observed vulnerability of deeper NIDS models can be attributed to increased model complexity in a constrained feature space. Research shows that deeper networks exhibit higher complexity, which has an inverse relationship with adversarial robustness \cite{khachaturov2024complexity}. In the NIDS domain with its low-dimensional input, each additional layer substantially increases model complexity relative to the limited input dimensions, \textbf{amplifying sensitivity to gradient-based perturbations.}

\item Additionally, the heterogeneous and structured nature of NIDS features may contribute to the observed results, as deeper networks can learn more complex, but potentially exploitable, representations. These domain-specific architectural effects highlight the need for tailored approaches in both model design and defense strategies for ML-NIDS compared to image processing tasks.

\item In a nutshell, unlike the computer vision domain (already complex and high-dimensional) where \(\varepsilon\) is the most influential parameter for robustness, in the NIDS domain, it is more complicated, and in addition to \(\varepsilon\), model parameters and neural network depth play important roles.

\end{itemize}

\section{Discussion and Implications}

Our study highlights key domain-specific differences in developing neural networks that are robust against adversarial attacks. In the NIDS domain, increasing model depth can unintentionally make models more susceptible to adversarial perturbations, as a larger parameter space and complex decision boundaries expand the attack surface. This finding challenges a common practice in machine learning, using deeper networks for better performance, by demonstrating that, in the context of security-sensitive applications, added complexity can backfire unless carefully controlled. Therefore, shallow models and dimensionality reduction may serve as effective first-line defenses in NIDS tasks. Conversely, in high-dimensional, structured domains like image recognition (e.g., MNIST), the effect of depth is minimal, and adversarial effectiveness depends mainly on the perturbation magnitude (epsilon). These results imply that effective adversarial defense strategies must be tailored to the specific feature space and domain, rather than universally applying approaches that work in computer vision to network security.

Furthermore, our findings highlight the importance of feature engineering and model design for security-critical systems \cite{elshehaby2024introducing}. Practitioners should not only focus on accuracy but also consider robustness measures and attack surface minimization during the model selection and deployment phases. This work encourages future research into defense strategies specialized for heterogeneous and structured data, such as NIDS, where traditional deep learning intuitions may not hold.




\section{Conclusion}

This paper explores the relationship between neural network depth and adversarial robustness in both network intrusion detection (CSE-CIC-IDS2018) and image processing (MNIST) domains. Our key finding is that, while increasing depth does not significantly affect adversarial robustness in image classification tasks, it can actually reduce robustness in network intrusion detection, making deeper NIDS models more vulnerable to adversarial attacks. These results demonstrate that adding parameters to neural networks is not always better for security, and underscore the necessity for domain-aware architecture choices when defending against evasion attacks. Future work should investigate tailored defenses that balance predictive accuracy and adversarial resilience, especially in structured-data settings like NIDS.




\section{Acknowledgment}
 This work was supported by the Natural Sciences and Engineering Research Council of Canada (NSERC) through the NSERC Discovery Grant program.

\bibliographystyle{IEEEtran}
\bibliography{main}

\end{document}